\documentclass[runningheads]{llncs}
\usepackage{graphicx}

\usepackage{cite}
\usepackage{amsmath,amssymb,amsfonts}
\usepackage{algorithmic}
\usepackage{graphicx}
\usepackage{textcomp}
\usepackage{xcolor,colortbl}
\def\BibTeX{{\rm B\kern-.05em{\sc i\kern-.025em b}\kern-.08em
    T\kern-.1667em\lower.7ex\hbox{E}\kern-.125emX}}
    
\usepackage{mathtools}    
\usepackage{colortbl}
\usepackage{url}    
\usepackage[inline]{enumitem}
\usepackage{tikz}
\usetikzlibrary{fit}
\usetikzlibrary{arrows.meta}
\usepackage{eqparbox}
\newbox\nodebox
\usetikzlibrary{trees,shapes.geometric,calc}
\usetikzlibrary{positioning}
\pgfdeclarelayer{back}
\pgfsetlayers{back,main}

\usepackage{caption}
\usepackage{subcaption}

\usepackage{tabu}
\usepackage[inline]{enumitem}

\definecolor{logMove}{HTML}{000000}
\definecolor{modelMove}{HTML}{666666}
\definecolor{synchronousMove}{HTML}{d9d9d9}

\usepackage[inline]{enumitem}

\usepackage{makecell}
\usepackage{tabularx}
\usepackage{multicol}
\colorlet{comment}{blue}

\usepackage[ruled,lined,vlined]{algorithm2e}

\SetCommentSty{xCommentSty}

\usepackage{hyperref}

\newcommand{\pt}{T{=}(V,\allowbreak E,\allowbreak \lambda,\allowbreak r)}
\newcommand{\bool}{\ensuremath{\{0,1\}}}
\newcommand{\emptyseq}{\ensuremath{\langle\rangle}}

\usepackage[nameinlink,capitalize]{cleveref}

\makeatletter
\g@addto@macro\normalsize{%
  \setlength\abovedisplayskip{2pt}
  \setlength\belowdisplayskip{2pt}
  \setlength\abovedisplayshortskip{2pt}
  \setlength\belowdisplayshortskip{2pt}
}
\makeatother

\begin{document}
\title{Alignment Approximation for Process Trees}
%
%
\author{Daniel Schuster\inst{1} \and
Sebastiaan van Zelst\inst{1,2} \and
Wil M. P. van der Aalst\inst{1,2}}
\authorrunning{D. Schuster et al.}
%
\institute{Fraunhofer Institute for Applied Information Technology FIT, Germany\\
\email{\{daniel.schuster,sebastiaan.van.zelst\}@fit.fraunhofer.de}
\and
RWTH Aachen University, Germany\\
\email{wvdaalst@pads.rwth-aachen.de}}
\maketitle              
\begin{abstract}
Comparing observed behavior (event data generated during process executions) with modeled behavior (process models), is an essential step in process mining analyses. 
Alignments are the de-facto standard technique for calculating conformance checking statistics.
However, the calculation of alignments is computationally complex since a shortest path problem must be solved on a state space which grows non-linearly with the size of the model and the observed behavior, leading to the well-known \emph{state space explosion problem}.
In this paper, we present a novel framework to approximate alignments on process trees by exploiting their hierarchical structure.
Process trees are an important process model formalism used by state-of-the-art process mining techniques such as the inductive mining approaches.
Our approach exploits structural properties of a given process tree and splits the alignment computation problem into smaller sub-problems.
Finally, sub-results are composed to obtain an alignment. 
Our experiments show that our approach provides a good balance between accuracy and computation time.

\end{abstract}
\section{Introduction}
\emph{Conformance checking} is a key research area within process mining~\cite{DBLP:books/sp/Aalst16}.
The comparison of observed process behavior with reference process models is of crucial importance in process mining use cases.
Nowadays, \emph{alignments}~\cite{DBLP:journals/widm/AalstAD12} are the de-facto standard technique to compute conformance checking statistics.
However, the computation of alignments is complex since a shortest path problem must be solved on a non-linear state space composed of the reference model and the observed process behavior.
This is known as the \emph{state space explosion problem}~\cite{DBLP:books/sp/CarmonaDSW18}.
Hence, various approximation techniques have been introduced.
Most techniques focus on decomposing Petri nets or reducing the number of alignments to be calculated when several need to be calculated for the same process model~\cite{DBLP:journals/isci/LeeVMAS18,DBLP:conf/caise/SaniZA20,DBLP:conf/bpm/TaymouriC18,DBLP:conf/simpda/TaymouriC16,DBLP:conf/bpm/0006AW19}.

In this paper, we focus on a specific class of process models, namely process trees (also called \emph{block-structured} process models), which are an important process model formalism that represent a subclass of sound \emph{Workflow nets}\cite{leemans-phd}.
For instance, various state-of-the-art process discovery algorithms return process trees~\cite{leemans-phd,DBLP:conf/bpm/LeemansFA13,schuster_incr_process_discovery}.
In this paper, we introduce an alignment approximation approach for process trees that consists of two main phases.
First, our approach splits the problem of alignments into smaller sub-problems along the tree hierarchy.
Thereby, we exploit the hierarchical structure of process trees and their semantics.
Moreover, the definition of sub-problems is based on a \emph{gray-box view} on the corresponding subtrees since we use a simplified/abstract view on the subtrees to recursively define the sub-problems along the tree hierarchy.
Such sub-problems can then be solved individually and in parallel.
Secondly, we recursively compose an alignment from the sub-results for the given process tree and observed process behavior.
Our experiments show that our approach provides a good balance between accuracy and computation effort.


The remainder is structured as follows. 
In \autoref{sec:related_work}, we present related work.
In \autoref{sec:preliminaries}, we present preliminaries.
In \autoref{sec:formal_framework}, we present the formal framework of our approach.
In \autoref{sec:approx_algo}, we introduce our alignment approximation approach.
In \autoref{sec:evaluation}, we present an evaluation.
\autoref{sec:conclusion} concludes the paper.

\section{Related Work}
\label{sec:related_work}
In this section, we present related work regarding alignment computation and approximation.
For a general overview of conformance checking, we refer to \cite{DBLP:books/sp/CarmonaDSW18}.

Alignments have been introduced in~\cite{DBLP:journals/widm/AalstAD12}.
In~\cite{adriansyah_2014_phd_aligning} it was shown that the computation is reducible to a shortest path problem and the solution of the problem using the A* algorithm is presented. 
In~\cite{DBLP:conf/bpm/Dongen18}, the authors present an improved heuristic that is used in the shortest path search.
In~\cite{DBLP:conf/caise/DongenCCT17}, an alignment approximation approach based on approximating the shortest path is presented. 

A generic approach to decompose Petri nets into multiple sub-nets is introduced in~\cite{DBLP:journals/dpd/Aalst13}.
Further, the application of such decomposition to alignment computation is presented.
In contrast to our approach, the technique does not return an alignment.
Instead, only partial alignments are calculated, which are used, for example, to approximate an overall fitness value.  
In~\cite{DBLP:journals/isci/LeeVMAS18}, an approach to calculate alignments based on Petri net decomposition~\cite{DBLP:journals/dpd/Aalst13} is presented that additionally guarantees optimal fitness values and optionally returns an alignment.
Comparing both decomposition techniques with our approach, we do not calculate sub-nets because we simply use the given hierarchical structure of a process tree.
Moreover, our approach always returns a valid alignment.

In~\cite{DBLP:conf/caise/SaniZA20}, an approach is presented that approximates alignments for an event log by reducing the number of alignments being calculated based on event log sampling.
Another technique based on event log sampling is presented in~\cite{DBLP:conf/bpm/0006AW19} where the authors explicitly approximate conformance results, e.g., fitness, rather than alignments.
In contrast to our proposed approach, alignments are not returned.
In~\cite{DBLP:conf/bpm/TaymouriC18} the authors present an approximation approach that explicitly focuses on approximating multiple optimal alignments.
Finally, in~\cite{DBLP:conf/simpda/TaymouriC16}, the authors present a technique to reduce a given process model and an event log s.t. the original behavior of both is preserved as much as possible.
In contrast, the proposed approach in this paper does not modify the given process model and event log.

\section{Preliminaries}
\label{sec:preliminaries}


We denote the power set of a given set $X$ by $\mathcal{P}(X)$.
A multi-set over a set $X$ allows multiple appearances of the same element.
We denote the universe of multi-sets for a set $X$ by $\mathcal{B}(X)$ and the set of all sequences over $X$ as $X^*$, e.g., $\langle a,b,b \rangle{\in} \{a,b,c\}^*$. 
For a given sequence $\sigma$, we denote its length by $|\sigma|$. 
We denote the empty sequence by $\emptyseq$.
We denote the set of all possible permutations for given $\sigma{\in}X^*$ by $\mathbb{P}(\sigma){\subseteq}X^*$.
Given two sequences $\sigma$ and $\sigma'$, we denote the concatenation of these two sequences by $\sigma {\cdot} \sigma'$. 
We extend the $\cdot$ operator to sets of sequences, i.e., let $S_1,S_2{\subseteq}X^*$ then $S_1{\cdot}S_2{=}\{\sigma_1{\cdot}\sigma_2\ {|} \sigma_1{\in}S_1 {\land}\allowbreak \sigma_2{\in}S_2\}$.
For traces $\sigma,\sigma'$, the set of all interleaved sequences is denoted by $\sigma {\diamond} \sigma'$, e.g., $\langle a,b \rangle {\diamond} \langle c \rangle {=} \{ \langle a,b,c\rangle,\allowbreak\langle a,c,b\rangle,\allowbreak\langle c,a,b\rangle \}$.
We extend the $\diamond$ operator to sets of sequences. 
Let $S_1,S_2 {\subseteq} X^*$, $S_1{\diamond} S_2$ denotes the set of interleaved sequences, i.e., $S_1{\diamond} S_2 {=}\allowbreak {\bigcup}_{\sigma_1{\in}S_1,\sigma_2{\in}S_2} \sigma_1 {\diamond} \sigma_2$.


For $\sigma{\in}X^*$ and $X'{\subseteq}{X}$, we recursively define the projection function $\sigma_{\downarrow_{X'}} {:} X^* {\to}\allowbreak (X')^*$ with: $\emptyseq_{\downarrow_{X'}} {=} \emptyseq$,\allowbreak $\big(\langle x\rangle {\cdot} \sigma\big)_{\downarrow_{X'}} {=} \langle x\rangle {\cdot} \sigma_{\downarrow_{X'}}$ if $x{\in}X'$ and $(\langle x\rangle {\cdot} \sigma)_{\downarrow_{X'}} {=} \sigma_{\downarrow_{X'}}$ else.

Let $t{=}(x_1,\dots,x_n){\in}X_1{\times}  \dots {\times} X_n$ be an $n$-tuple over $n$ sets.
We define projection functions that extract a specific element of $t$, i.e., $\pi_1(t){=}x_1,\dots, \pi_n(t){=}x_n$, e.g., $\pi_2\left(\left( a,b,c \right)\right){=}b$. 
Analogously, given a sequence of length $m$ with $n$-tuples $\sigma{=}\langle(x^1_1,\dots,x^1_n),\dots, \allowbreak (x_1^m,\dots,x_n^m)\rangle$, we define $\pi^*_1(\sigma){=}\langle x_1^1,\dots,\allowbreak x_1^m\rangle, \dots,\allowbreak\pi^*_n(\sigma){=}\allowbreak\langle x_n^1,\dots,x_n^m\rangle$.
For instance,  $\pi^*_2\big(\langle (a,b), (a,c), (b,a) \rangle\big) {=} \langle b,c,a \rangle$.

\subsection{Event Logs}

Process executions leave \emph{event data} in information systems.
An \emph{event} describes the execution of an activity for a particular \emph{case}/process instance.
Consider \autoref{tab:event_log} for an example of an \emph{event log} where each event contains the executed activity, a timestamp, a case-id and potentially further attributes.
Since, in this paper, we are only interested in the sequence of activities executed, we define an event log as a multi-set of sequences.
Such sequence is also referred to as a \emph{trace}.

\begin{definition}[Event log]
Let $\mathcal{A}$ be the universe of activities. 
$L{\in}\mathcal{B}(\mathcal{A}^*)$ is an event log. 
\end{definition}

\begin{table}[tb]
    \caption{Example of an event log from an order process}
    \label{tab:event_log}
    \centering
    \scriptsize
    \begin{tabular}{ c c c c c }
        \hline
         \textbf{Event-id} & \textbf{Case-id} & \textbf{Activity name} & \textbf{Timestamp} & \textbf{$\cdots$}  \\ \hline
         $\cdots$ & $\cdots$ & $\cdots$ & $\cdots$ & $\cdots$ \\
         
         $200$ & $13$ & create order (c) & 2020-01-02 15:29  & $\cdots$ \\
         $201$ & $27$ & receive payment (r) & 2020-01-02 15:44 & $\cdots$ \\
         $202$ & $43$ & dispatch order (d) & 2020-01-02 16:29 & $\cdots$ \\
         $203$ & $13$ & pack order (p) & 2020-01-02 19:12 & $\cdots$ \\
         
         $\cdots$ & $\cdots$ & $\cdots$ & $\cdots$ & $\cdots$ \\ \hline
    \end{tabular}
\end{table}

\subsection{Process Trees}

Next, we define the syntax and semantics of process trees.

\begin{definition}[Process Tree Syntax]
\label{def:process-tree}
Let $\mathcal{A}$ be the universe of activities and $\tau {\notin}  \mathcal{A}$. 
Let $\bigoplus {=} \{\rightarrow,\times,\wedge,\circlearrowleft\}$ be the set of process tree operators.
We define a process tree $T{=}(V,E,\lambda,r)$ consisting of a totally ordered set of nodes $V$, a set of edges $E$, a labeling function $\lambda {:} V {\to} \mathcal{A}{\cup}\{\tau\}{\cup}\bigoplus$ and a root node $r{\in}V$. 

\begin{itemize}[noitemsep,topsep=0pt]
    \item $\big( \{n\},\{\},\lambda,n \big)$ with $\lambda(n){\in} \mathcal{A}{\cup}\{\tau\}$ is a process tree

   \item given $k{>}1$ process trees $T_1{=}(V_1,E_1,\lambda_1,r_1),\dots, \allowbreak T_k{=}\allowbreak(V_k,\allowbreak E_k,\allowbreak\lambda_k,r_k)$, $\pt$ is a process tree s.t.:
   \begin{itemize}[noitemsep,topsep=0pt]
       \item $V{=}V_1{\cup}\dots{\cup}V_k{\cup}\{r\}$ (assume $r{\notin} V_1{\cup}\dots{\cup}V_k$)
       \item $E{=}E_1{\cup}\dots{\cup}E_k{\cup}\big\{ (r,r_1),\dots,(r,r_k) \big\}$
       \item $\lambda(x){=}\lambda_j(x) \ \forall j{\in}\{1,\dots,k\}  \forall x{\in}V_j, \lambda(r){\in}\{\rightarrow,\wedge,\times\}$
   \end{itemize}
   
   \item given two process trees $T_1{=}(V_1,E_1,\lambda_1,r_1)$ and $T_2{=}\allowbreak(V_2,\allowbreak E_2,\allowbreak\lambda_2,\allowbreak r_2)$, $\pt$ is a process tree s.t.:
   \begin{itemize}[noitemsep,topsep=0pt]
       \item $V{=}V_1{\cup}V_2{\cup}\{r\}$ (assume $r{\notin} V_1{\cup}V_2$)
       \item $E{=}E_1{\cup}E_2{\cup}\big\{ (r,r_1),(r,r_2) \big\}$
       \item $\lambda(x){=}\lambda_1(x)$ if $x{\in}V_1, \lambda(x){=}\lambda_2(x)$ if $x{\in}V_2, \lambda(r){=}\circlearrowleft$
   \end{itemize}
\end{itemize}
\end{definition}

\begin{figure}[tb]
    \centering
    \footnotesize
    \resizebox{.65\textwidth}{!}{
    \begin{tikzpicture}[every label/.style={text=darkgray,font=\scriptsize}]
        \tikzstyle{tree_op}=[rectangle,draw=black,fill=gray!30,thick,minimum size=5mm,inner sep=0pt]
        \tikzstyle{tree_leaf}=[circle,draw=black,thick,minimum size=5mm,inner sep=0pt]
        \tikzstyle{tree_leaf_inv}=[circle,draw=black,fill=black,thick,minimum size=5mm, text=white,inner sep=0pt]
        \tikzstyle{marking}=[dashed, draw=lightgray]

        \tikzstyle{level 1}=[sibling distance=40mm,level distance=5mm]
        \tikzstyle{level 2}=[sibling distance=20mm,level distance=5mm]
        \tikzstyle{level 3}=[sibling distance=20mm, level distance= 5mm]
        \tikzstyle{level 4}=[sibling distance=12mm,level distance=5mm]
        
        \node [tree_op, label=$n_0$] (root){$\rightarrow $}
          child {node [tree_op, label=$n_{1.1}$] (loop) {$\circlearrowleft $}
            child { node [tree_op, label=$n_{2.1}$] (choice) {$\times$}
              child {node [tree_op, label=$n_{3.1}$] (sequence) {$\rightarrow$}
                child {node [tree_leaf, label=$n_{4.1}$] (a) {$a$}}
                child {node [tree_leaf, label=$n_{4.2}$] (b) {$b$}}
              } 
              child {node [tree_op,label=$n_{3.2}$] (parallel1) {$\wedge$}
                child {node [tree_leaf, label=$n_{4.3}$] (c) {$c$}}
                child {node [tree_leaf, label=$n_{4.4}$] (d) {$d$}}
              } 
            }
            child {node [tree_leaf_inv, label=$n_{2.2}$] (tau) {$\tau$}}
          }
          child {node [tree_op, label=$n_{1.2}$] (parallel2) {$\wedge$}
            child {node [tree_leaf, label=$n_{2.3}$] (e) {$e$}}
          	child {node [tree_leaf, label=$n_{2.4}$] (f) {$a$}}
          }
        ;
 
       \node[marking,fit=(loop) (a) (d) (tau) ,inner sep=5pt,yshift=0.1cm, label=above:{$T_1 {=} \triangle^{T_0}(n_{1.1})$}] {};
       \node[marking,fit=(parallel2) (e) (f) ,inner sep=5pt,yshift=0.1cm, label=above:{$T_2 {=} \triangle^{T_0}(n_{1.2})$}] {};
     
    \end{tikzpicture}
    }
    \caption{Process tree $T_0 {=}\big(  \{n_o,\dots,n_{4.4}\},\allowbreak \big\{ (n_0,n_{1.1}),\allowbreak\dots,\allowbreak (n_{3.2},n_{4.4})\big\},\allowbreak \lambda,\allowbreak n_0\big)$ with $\lambda(n_0){=}{\rightarrow},\dots,\allowbreak\lambda(n_{4.4}){=}d$}
    
    \label{fig:process_tree_example}
\end{figure}

In \autoref{fig:process_tree_example}, we depict an example process tree $T_0$ that can alternatively be represented textually due to the totally ordered node set, i.e., $T_0 {\widehat{=}} {\rightarrow}( {\circlearrowleft} ( {\times} ( {\rightarrow} (a,b) ,  \allowbreak{\wedge} (c,d)    ) ,\tau )   ,\allowbreak {\wedge} (e,a) )$.
We denote the universe of process trees by $\mathcal{T}$.
The degree $d$ indicates the number of edges connected to a node.
We distinguish between incoming $d^+$ and outgoing edges $d^-$, e.g., $d^+(n_{2.1}){=}1$ and $d^-(n_{2.1}){=}2$. 
For a tree $\pt$, we denote its \emph{leaf nodes} by $T^L{=}\{v{\in}V {\mid} d^-(v){=}0\}$.
The child function $c^T {:} V {\to} V^*$ returns a sequence of child nodes according to the order of $V$, i.e., $c^T(v){=}\langle v_1,\dots,v_j\rangle$ s.t. $(v,v_1),\dots,(v,v_j){\in}E$.
For instance, $c^T(n_{1.1}){=}\langle n_{2.1},\allowbreak n_{2.2}\rangle$. 
For $\pt$ and a node $v{\in}V$, $\triangle^T(v)$ returns the corresponding tree $T'$ s.t. $v$ is the root node, i.e., $T'{=}(V',E',\lambda',v)$.
Consider $T_0$, $\triangle^{T_0}(n_{1.1}){=}T_1$ as highlighted in \autoref{fig:process_tree_example}.
For process tree $T{\in}\mathcal{T}$, we denote its height by $h(T){\in}\mathbb{N}$. 

\begin{definition}[Process Tree Semantics]
For given $\pt\allowbreak{\in}\mathcal{T}$, we define its language $\mathcal{L}(T){\subseteq}\mathcal{A}^*$.
\begin{itemize}[noitemsep,topsep=0pt]
    \item if $\lambda(r){=}a{\in}\mathcal{A}$, $\mathcal{L}(T){=}\{\langle a \rangle\}$
    
    \item if $\lambda(r){=}\tau$, $\mathcal{L}(T){=}\{\langle  \rangle\}$
    
    \item if $\lambda(r){\in}\{\rightarrow,\times,\wedge\}$ with $c^T(r){=}\langle v_1,\dots,v_k\rangle$
    \begin{itemize}
        \item with $\lambda(r){=}{\rightarrow}$, $\mathcal{L}(T){=}\mathcal{L}(\triangle^T(v_1)) {\cdot}\dots{\cdot}  \mathcal{L}(\triangle^T(v_k))$
        
        \item with $\lambda(r){=}{\wedge}$, $\mathcal{L}(T){=}\mathcal{L}(\triangle^T(v_1)) {\diamond}\dots{\diamond} \mathcal{L}(\triangle^T(v_k))$ 
        
        \item with $\lambda(r){=}{\times}$, $\mathcal{L}(T){=}\mathcal{L}(\triangle^T(v_1)) {\cup}\dots{\cup}  \mathcal{L}(\triangle^T(v_k))$

    \end{itemize}
    \item if $\lambda(r){=}{\circlearrowleft}$ with $c^T(r){=}\langle v_1,v_2\rangle$, 
    $\mathcal{L}(T){=}\{ \sigma_1 {\cdot} \sigma_1' {\cdot} \sigma_2 {\cdot} \sigma_2' {\cdot}\allowbreak {\dots} {\cdot} \sigma_m \mid m {\geq} 1 \land \allowbreak {\forall}{ 1 {\leq} i {\leq} m}\allowbreak \big(\sigma_i{\in}\mathcal{L}(\triangle^T(v_1))\big) \land\allowbreak {\forall}{1 {\leq} i {\leq} m{-}1} \allowbreak \big(\sigma_i'{\in}\mathcal{L}(\triangle^T(v_2))\big) \}$ 
    
\end{itemize}
\end{definition}

In this paper, we assume binary process trees as input for our approach, i.e, every node has two or none child nodes, e.g., $T_0$.
Note that every process tree can be easily converted into a language equivalent binary process tree~\cite{leemans-phd}.

\subsection{Alignments}

\emph{Alignments}\cite{adriansyah_2014_phd_aligning} map observed behavior onto modeled behavior specified by process models.
\autoref{fig:alignments} visualizes an alignment for the trace $\langle a,b,c,f \rangle$ and $T_0$ (\autoref{fig:process_tree_example}). 
The first row corresponds to the given trace ignoring the skip symbol $\gg$.
The second row (ignoring $\gg$) corresponds to a sequence of leaf nodes s.t. the corresponding sequence of labels (ignoring $\tau$) is in the language of the process tree, i.e., $\langle a,b,d,c,a,e \rangle {\in} \mathcal{L}(T_0)$.
Each column represents an alignment move.
The first two are \emph{synchronous moves} since the activity and the leaf node label are equal.
The third and fourth are \emph{model moves} because $\gg$ is in the log part.
Moreover, the third is an \emph{invisible} model move since the leaf node label is $\tau$ and the fourth is a \emph{visible} model move since the label represents an activity. 
Visible model moves indicate that an activity should have taken place w.r.t. the model.
The sixth is a log move since the trace part contains $\gg$.
Log moves indicate observed behavior that should not occur w.r.t. the model.
Note that we alternatively write $\gamma {\widehat{=}} \big\langle (a,a),\dots,(\gg,e) \big\rangle$ using their labels instead of leaf nodes.

\begin{figure}[tb]
    \scriptsize
    \newcolumntype{l}{>{\columncolor{logMove}\color{white}}c}
    \newcolumntype{s}{>{\columncolor{synchronousMove}}c}
    \newcolumntype{m}{>{\columncolor{modelMove}\color{white}}c}
    \centering
    \begin{tabular}{|r||s | s | c | m | s | l | m | m |}\hline
        \color{gray}{\textbf{\emph{trace part}}} & $a$     & $b$     & $\gg$   & $\gg$ & $c$ & $f$ & $\gg$ & $\gg$   \\ \hline
        \color{gray}{\textbf{\emph{model part}}} & \makecell{$n_{4.1}$ \\ \color{gray}{$\lambda(n_{4.1}){=}a$}}     
        & \makecell{$n_{4.2}$ \\ \color{gray}{$\lambda(n_{4.2}){=}b$}}     
        & \makecell{$n_{2.2}$ \\ \color{gray}{$\lambda(n_{2.2}){=}\tau$}}  
        & \makecell{$n_{4.4}$ \\ \color{lightgray}{$\lambda(n_{4.4}){=}d$}}   
        & \makecell{$n_{4.3}$ \\ \color{gray}{$\lambda(n_{4.3}){=}c$}} 
        & \makecell{$\gg$} 
        & \makecell{$n_{2.4}$ \\ \color{lightgray}{$\lambda(n_{2.4}){=}a$}}
        & \makecell{$n_{2.3}$ \\ \color{lightgray}{$\lambda(n_{2.3}){=}e$}} \\ \hline
    \end{tabular}
    
    \caption{Optimal alignment $\gamma{=}\big\langle (a,n_{4.1}),\dots,(\gg,n_{2.3}) \big\rangle$ for $\langle a,b,c,f \rangle$ and $T_0$}
    \label{fig:alignments}
\end{figure}

\begin{definition}[Alignment]
Let $\mathcal{A}$ be the universe of activities, $\sigma {\in} \mathcal{A}^*$ be a trace and $\pt {\in} \mathcal{T}$ be a process tree with leaf nodes $T^L$. 
Note that $\gg,\tau {\notin} \mathcal{A}$.
A sequence $\gamma {\in} \big( (\mathcal{A}{\cup}\{\gg\})\allowbreak {\times}\allowbreak (T^L{\cup}\{\gg\} ) \big)^*$ with length $n{=}|\gamma|$ is an alignment iff:
\begin{enumerate}[noitemsep,topsep=0pt]
    \item $\sigma {=} \pi^*_1(\gamma)_{\downarrow_{\mathcal{A}}}$
    
    \item $\Big\langle \lambda \Big( \pi_2 \big( \gamma \left( 1\right) \big) \Big) , \dots , \lambda \Big( \pi_2 \big( \gamma(n) \big) \Big) \Big \rangle_{\downarrow_{\mathcal{A}}}  {\in} \mathcal{L}(T)$
    
    \item $(\gg,\gg) {\notin} \gamma$ and $(a,v) {\notin} \gamma \ \forall a {\in} \mathcal{A} \ \forall v {\in} T^L \big(a {\neq} \lambda(v)\big)$
\end{enumerate}
\end{definition}

For a given process tree and a trace, many alignments exist.
Thus, costs are assigned to alignment moves.
In this paper, we assume the \emph{standard cost function}.
Synchronous and invisible model moves are assigned cost 0, other moves are assigned cost 1. 
An alignment with minimal costs is called \emph{optimal}.
For a process tree $T$ and a trace $\sigma$, we denote the set of all possible alignments by $\Gamma(\sigma,T)$.
In this paper, we assume a function $\alpha$ that returns for given $T {\in}\mathcal{T}$ and $\sigma {\in} \mathcal{A}^*$ an optimal alignment, i.e., ${\alpha}(\sigma,T){\in} \Gamma(\sigma,T)$.
Since process trees can be easily converted into Petri nets~\cite{DBLP:books/sp/Aalst16} and the computation of alignments for a Petri net was shown to be reducible to a shortest path problem~\cite{adriansyah_2014_phd_aligning}, such function exists.

\section{Formal Framework}
\label{sec:formal_framework}

In this section, we present a general framework that serves as the basis for the proposed approach.
The core idea is to recursively divide the problem of alignment calculation into multiple sub-problems along the tree hierarchy. 
Subsequently, we recursively compose partial sub-results to an alignment.

Given a trace and tree, we recursively split the trace into sub-traces and assign these to subtrees along the tree hierarchy.
During splitting/assigning, we regard the semantics of the current root node's operator.
We recursively split until we can no longer split, e.g., we hit a leaf node.
Once we stop splitting, we calculate optimal alignments for the defined sub-traces on the assigned subtrees, i.e., we obtain sub-alignments. 
Next, we recursively compose the sub-alignments to a single alignment for the parent subtree.
Thereby, we consider the semantics of the current root process tree operator.
Finally, we obtain a \emph{valid}, but not necessarily optimal, alignment for the initial given tree and trace since we regard the semantics of the process tree during splitting/assigning and composing.

Formally, we can express the splitting/assigning as a function.
Given a trace $\sigma {\in} \mathcal{A}^*$ and $\pt {\in} \mathcal{T}$ with subtrees $T_1$ and $T_2$, $\psi$ splits the trace $\sigma$ into $k$ sub-traces $\sigma_1,\dots,\sigma_k$ and assigns each sub-trace to either $T_1$ or $T_2$.
\begin{equation}
\label{eq:splitting}
    \psi(\sigma,T) {\in} \Big\{ \big\langle (\sigma_1,T_{i_1}), \dots,  (\sigma_k,T_{i_k})\big\rangle \mid i_1,\dots,i_k{\in}\{1,2\} \land \sigma_1 {\cdot} \dots{\cdot}\sigma_k {\in}\mathbb{P}(\sigma) \Big\}
\end{equation}
We call a splitting/assignment \emph{valid} if the following additional conditions are satisfied depending on the process tree operator:
\begin{itemize}[noitemsep,topsep=0pt]
    \item if $\lambda(r) {=} {\times}$: $k{=}1$
    \item if $\lambda(r) {=} {\rightarrow}$: $k{=}2 \land \sigma_1{\cdot}\sigma_2{=}\sigma$
    \item if $\lambda(r) {=} {\wedge}$: $k{=}2$
    \item if $\lambda(r) {=} {\circlearrowleft}$: $k{\in}\{1,3,5,\dots\} \land \sigma_1{\cdot}\dots{\cdot}\sigma_k {=} \sigma \land i_1{=}1 \land \allowbreak  \forall j{\in} \{1,\dots,k{-}1\} \big( (i_j{=}1 {\Rightarrow}\allowbreak i_{j+1}{=}2) \land (i_j{=}2 {\Rightarrow}\allowbreak i_{j+1}{=}1) \big)$ 
\end{itemize}

Secondly, the calculated sub-alignments are recursively composed to an alignment for the respective parent tree.
Assume a tree $T {\in} \mathcal{T}$ with sub-trees $T_1$ and $T_2$, a trace $\sigma {\in} \mathcal{A}^*$, a valid splitting/assignment $\psi(\sigma,T)$ , and a sequence of $k$ sub-alignments $\langle \gamma_1,\dots,\gamma_k \rangle$ s.t. $\gamma_j{\in}\Gamma(\sigma_j,T_{i_j})$ with $(\sigma_j,T_{i_j}){=}\psi(\sigma,T)(j)  \forall j {\in} \{1,\dots,k\}$. 
The function $\omega$ composes an alignment for $T$ and $\sigma$ from the given sub-alignments.
\begin{equation}
\label{eq:composing}
    \omega(\sigma,T, \langle \gamma_1,\dots,\gamma_k \rangle) {\in} \{\gamma \mid \gamma {\in}\Gamma(\sigma,T) \land \gamma_1{\cdot} \dots {\cdot} \gamma_k{\in}\mathbb{P}(\gamma)\}
\end{equation}
By utilizing the definition of process tree semantics, it is easy to show that, given a valid splitting/assignment, such alignment $\gamma$ returned by $\omega$ always exists. 

\begin{algorithm}[tb]
    \scriptsize
	\caption{Approximate alignment}
	\label{alg:alignment_approx}
	\SetKwInOut{Input}{input}
	\Input{$\pt{\in}\mathcal{T}, \sigma{\in}\mathcal{A}^*, TL{\geq}1, TH{\geq}1$}
	\Begin{
		\nl \If{$ |\sigma| {\leq} TL \lor h(T) {\leq}TH $ }{\label{alg:line:tl_th}
		    \nl\Return ${\alpha}(\sigma,T)$\tcp*[r]{optimal alignment}
		}\label{alg:line:optimal}
		\nl \Else{
		    
		    \nl $\psi(\sigma,T) {=}\langle (\sigma_1,T_{i_1}), \dots,  (\sigma_k,T_{i_k})\big\rangle $\tcp*[r]{valid splitting}\label{alg:line:splitting}

		    \nl \For{$(\sigma_j,T_{i_j}) {\in} \big\langle (\sigma_1,T_{i_1}), \dots,  (\sigma_k,T_{i_k})\big\rangle$}{
		    
		        \nl $ \gamma_j \gets $approx. alignment for $\sigma_j$ and $T_{i_j}$\tcp*[r]{recursion}\label{alg:line:recursive}
		    } 
		    \nl $\gamma \gets \omega(\sigma,T,\langle \gamma_1,\dots,\gamma_k \rangle)$\tcp*[r]{composing}\label{alg:line_composing}
		    \nl \Return $\gamma$\;
		}
	}			
\end{algorithm}

The overall, recursive approach is sketched in \autoref{alg:alignment_approx}.
For a given tree $T$ and trace $\sigma$, we create a valid splitting/assignment (\autoref{alg:line:splitting}).
Next, we recursively call the algorithm on the determined sub-traces and subtrees (\autoref{alg:line:recursive}). 
If given thresholds for trace length ($TL$) or tree height ($TH$) are reached, we stop splitting and return an optimal alignment (\autoref{alg:line:optimal}).
Hence, for the sub-traces created, we eventually obtain optimal sub-alignments, which we recursively compose to an alignment for the parent tree (\autoref{alg:line_composing}).
Finally, we obtain a valid, but not necessarily optimal, alignment for $T$ and $\sigma$.

\section{Alignment Approximation Approach}
\label{sec:approx_algo}
Here, we describe our proposed approach, which is based on the formal framework introduced. 
First, we present an overview.
Subsequently, we present specific strategies for splitting/assigning and composing for each process tree operator.

\subsection{Overview}
\label{sec:overview}

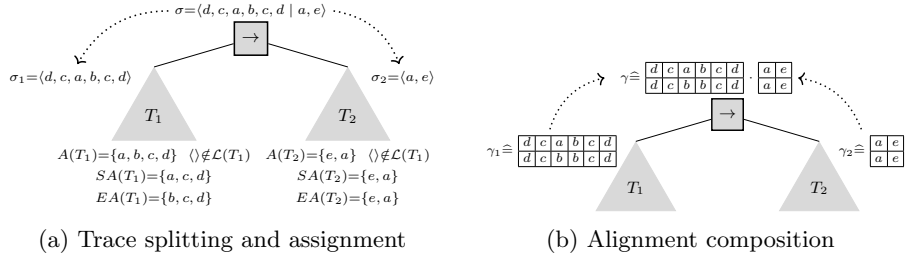
\begin{figure}[tb]
    \begin{subfigure}[b]{.49\columnwidth}
        \centering
        \resizebox{\textwidth}{!}{
            \begin{tikzpicture}
            \tikzset{
            tree_op/.style={rectangle,draw=black,fill=gray!30,thick,minimum size=6mm,inner sep=0pt},
            tree_subtree/.style={rectangle,draw = gray!15,fill=gray!15,thick,minimum size=6mm,inner sep=0pt},
            triangle/.style={inner sep=4pt,draw=gray!30, regular polygon, regular polygon sides=3, align=center,equal size=T,fill=gray!30},
            marking/.style={draw=gray},
            selected/.style={fill=darkgray, text=white, draw=gray},
            level 1/.style={sibling distance=35mm,level distance=14mm},
            equal size/.style={execute at begin
             node={\setbox\nodebox=\hbox\bgroup},
             execute at end
             node={\egroup\eqmakebox[#1][c]{\copy\nodebox}}}
            }

            \node [tree_op] (root){$\rightarrow $}
            [child anchor=north]
              child {node [triangle] (t1) {$T_1$}}
              child {node [triangle] (t2) {$T_2$}}
            ;
            \node [above of= root,yshift=-0.5cm] (sigma) {\scriptsize$\sigma{=}\langle d,c,a,b,c,d \; | \; a,e\rangle$};
            \node [below of= t1,yshift=-0.1cm,xshift=0cm,align=center] (label_t1) {\scriptsize$A(T_1){=}\{a,b,c,d\}$
            \scriptsize $\ \emptyseq{\notin}\mathcal{L}(T_1)$
            \\\scriptsize $SA(T_1){=}\{a,c,d\}$
            \\\scriptsize $EA(T_1){=}\{b,c,d\}$};
            
             \node [below of= t2,yshift=-.1cm,xshift=0cm,align=center] (label_t1) {\scriptsize$A(T_2){=}\{e,a\}$
             \scriptsize $\ \emptyseq{\notin}\mathcal{L}(T_1)$
            \\\scriptsize $SA(T_2){=}\{e,a\}$
            \\\scriptsize $EA(T_2){=}\{e,a\}$};
            
            \node [left of= t1,yshift=0.7cm,xshift=-.5cm] (sigma1) {\scriptsize$\sigma_1{=}\langle d,c,a,b,c,d\rangle$};
            \node [right of= t2,yshift=0.7cm,xshift=0cm] (sigma2) {\scriptsize$\sigma_2{=}\langle a,e\rangle$};
            
            \draw[->,bend right, dotted,thick](sigma.west) to (sigma1);
            \draw[->,bend left, dotted,thick](sigma.east) to (sigma2);

        \end{tikzpicture}
        }
        \caption{Trace splitting and assignment} 
        \label{fig:trace_splitting}
    \end{subfigure}
    \hfill
    \begin{subfigure}[b]{.49\columnwidth}
        \centering
            \resizebox{\textwidth}{!}{
            \begin{tikzpicture}
            \tikzset{
            tree_op/.style={rectangle,draw=black,fill=gray!30,thick,minimum size=6mm,inner sep=0pt},
            tree_subtree/.style={rectangle,draw = gray!15,fill=gray!15,thick,minimum size=6mm,inner sep=0pt},
            triangle/.style={inner sep=4pt,draw=gray!30, regular polygon, regular polygon sides=3, align=center,equal size=T,fill=gray!30},
            marking/.style={draw=gray},
            selected/.style={fill=darkgray, text=white, draw=gray},
            level 1/.style={sibling distance=35mm,level distance=14mm},
            equal size/.style={execute at begin
             node={\setbox\nodebox=\hbox\bgroup},
             execute at end
             node={\egroup\eqmakebox[#1][c]{\copy\nodebox}}}
            }

            \node [tree_op] (2root) {$\rightarrow $}
            [child anchor=north]
              child {node [triangle] (t1_) {$T_1$}}
              child {node [triangle] (t2_) {$T_2$}}
            ;
            
            \node [left of= t1_,yshift=0.7cm,xshift=-.7cm] (align_1) {
            \scriptsize
            \bgroup
            \setlength\tabcolsep{.08cm}
            \def\arraystretch{1}
            $\gamma_1{\widehat{=} }$
            \begin{tabular}{|c | c | c | c | c | c | c | c |}\hline
                    $d$     & $c$     & $a$   & $b$ & $c$ & $d$   \\ \hline
                    $d$     & $c$     & $b$   & $b$ & $c$ & $d$  \\ \hline
                \end{tabular}
            \egroup
            };
            
            \node [right of= t2_,yshift=0.7cm,xshift=-.1cm] (align_2) {
            \scriptsize
            \bgroup
            \setlength\tabcolsep{.08cm}
            \def\arraystretch{1}
            $\gamma_2{\widehat{=} }$
            \begin{tabular}{|c | c | c | c | c | c | c | c |}\hline
                    $a$     & $e$   \\ \hline
                    $a$     & $e$   \\ \hline
                \end{tabular}
            \egroup
            };
            
            \node [above of= 2root,yshift=-.3cm,xshift=-.5cm] (align) {
            \scriptsize
            \bgroup
            \setlength\tabcolsep{.08cm}
            \def\arraystretch{1}
            $\gamma{\widehat{=} }$
            \begin{tabular}{|c | c | c | c | c | c | c | c |c |c |}\hline
                    $d$     & $c$     & $a$   & $b$ & $c$ & $d$   \\ \hline
                    $d$     & $c$     & $b$   & $b$ & $c$ & $d$ \\ \hline
                \end{tabular}
                ${\cdot}$
                \begin{tabular}{|c | c | c | c | c | c | c | c |c |c |}\hline
                     $a$ & $e$   \\ \hline
                     $a$ & $e$ \\ \hline
                \end{tabular}
            \egroup
            };
            \draw[->,bend left, dotted,thick](align_1) to (align.west);
            \draw[->,bend right, dotted,thick](align_2) to (align.east);

        \end{tikzpicture}
        }
        \caption{Alignment composition} 
        \label{fig:alignment_composition}
    
    \end{subfigure}
     \caption{Overview of the two main actions of the approximation approach } 
    \label{fig:overview}
\end{figure}


For splitting a trace and assigning sub-traces to subtrees many options exist.
Moreover, it is inefficient to try out all possible options. 
Hence, we use a \emph{heuristic} that guides the splitting/assigning.
For each subtree, we calculate four characteristics: the activity labels $A$, if the empty trace is in the subtree's language, possible start-activities $SA$ and end-activities $EA$ of traces in the subtree's language. 
Thus, each subtree is a \emph{gray-box} since only limited information is available.

Consider the trace to be aligned $\sigma{=}\langle d,c,a,b,c,d,a,e \rangle$ and the two subtrees of $T_0$ with corresponding characteristics depicted in \autoref{fig:trace_splitting}.
Since $T_0$'s root node is a sequence operator, we need to split $\sigma$ once to obtain two sub-traces according to the semantics.
Thus, we have $9$ potential splittings positions:
$\langle \textcolor{gray}{\mid_1} \;d\; \textcolor{gray}{\mid_2} \;c\; \textcolor{gray}{\mid_3} \;a\; \textcolor{gray}{\mid_4} \;b\; \textcolor{gray}{\mid_5} \; c \allowbreak\; \textcolor{gray}{\mid_6} \;d\;\allowbreak \textcolor{gray}{\mid_7} \;a\; \textcolor{gray}{\mid_8} \;e\; \textcolor{gray}{\mid_9} \rangle$.
If we split at position $1$, we assign $\sigma_1{=}\emptyseq$ to the first subtree $T_1$ and the remaining trace $\sigma_2{=}\sigma$ to $T_2$.
Certainly, this is not a good decision since we know that $\emptyseq{\notin}\mathcal{L}(T_1)$, the first activity of $\sigma_2$ is not a start activity of $T_2$ and the activities $b,c,d$ occurring in $\sigma_2$ are not in $T_2$.

Assume we split at position $7$ (\autoref{fig:trace_splitting}).
Then we assign $\sigma_1{=}\langle d,c,a,b,c,d \rangle$ to $T_1$.
All activities in $\sigma_1$ are contained in $T_1$, $\sigma_1$ starts with $d{\in}SA(T_1)$ and ends with $d{\in}EA(T_1)$.
Further, we obtain $\sigma_2{=}\langle a,e \rangle$ whose activities can be replayed in $T_2$, and start- and end-activities match, too.
Hence, according to the gray-box-view, splitting at position 7 is a good choice.
Next, assume we receive two alignments $\gamma_1$ for $T_1,\sigma_1$  and $\gamma_2$ for $T_2,\sigma_2$ (\autoref{fig:alignment_composition}).
Since $T_1$ is executed before $T_2$, we concatenate the sub-alignments $\gamma {=} \gamma_1 {\cdot} \gamma_2$ and obtain an alignment for $T_0$.

\subsection{Calculation of Process Tree Characteristics}
In this section, we formally define the computation of the four tree characteristics for a given process tree $\pt$.
We define the activity set $A$ as a function, i.e., $A{:}\mathcal{T} {\to}  \mathcal{P}(\mathcal{A})$, with $A(T) {=} \{ \lambda(n) \mid n{\in} T^L, \lambda(n){\neq}\tau\}$. 
We recursively define the possible start- and end-activities as a function, i.e., $SA {:} \mathcal{T} {\to} \mathcal{P}(\mathcal{A})$ and $EA {:} \mathcal{T} {\to} \mathcal{P}(\mathcal{A})$. 
If $T$ is not a leaf node, we refer to its two subtrees as $T_1$ and $T_2$.

\scriptsize
\noindent\parbox[t]{.5\textwidth}{
\resizebox{.5\textwidth }{!}{ 
$
SA(T){=}
\begin{cases*}
      \{\lambda(r)\} & if $\lambda(r) {\in} \mathcal{A}$ \\
      \emptyset & if $\lambda(r) {=} \tau$ \\
      SA(T_1) & if $\lambda(r){=}{\rightarrow} {\land} \emptyseq{\notin} \mathcal{L}(T_1) $\\
      SA(T_1){\cup}SA(T_2) & if $\lambda(r){=}{\rightarrow} {\land} \emptyseq{\in}\mathcal{L}(T_1) $\\
      SA(T_1){\cup}SA(T_2) & if $\lambda(r){\in}\{\wedge,\times\} $\\
      SA(T_1) & if $\lambda(r){=}{\circlearrowleft} {\land} \emptyseq{\notin}\mathcal{L}(T_1) $\\
      SA(T_1){\cup}SA(T_2) & if $\lambda(r){=}{\circlearrowleft} {\land} \emptyseq{\in}\mathcal{L}(T_1) $
\end{cases*}
$
}
}
\hfill
\parbox[t]{.5\textwidth}{
\resizebox{.5\textwidth }{!}{ 
$
EA(T){=}
\begin{cases*}
     \{\lambda(n)\} & if $\lambda(r) {\in} \mathcal{A}$ \\
     \emptyset & if $\lambda(r) {=} \tau$ \\
      EA(T_2) & if $\lambda(r){=}{\rightarrow} {\land} \emptyseq{\notin} \mathcal{L}(T_2) $\\
      EA(T_1){\cup}EA(T_2) & if $\lambda(r){=}{\rightarrow} {\land} \emptyseq{\in}\mathcal{L}(T_2) $\\
      EA(T_1){\cup}EA(T_2) & if $\lambda(r){\in}\{\wedge,\times\}$\\
      EA(T_1) & if $\lambda(r){=}{\circlearrowleft} {\land} \emptyseq{\notin}\mathcal{L}(T_1) $\\
      EA(T_1){\cup}EA(T_2) & if $\lambda(r){=}{\circlearrowleft} {\land} \emptyseq{\in}\mathcal{L}(T_1) $ 
\end{cases*}
$
}
}
\normalsize

\noindent The calculation whether the empty trace is accepted can also be done recursively. 
\begin{itemize}[noitemsep,topsep=1pt]
    \item $\lambda(r){=}\tau \Rightarrow \emptyseq{\in}\mathcal{L}(T)$ and $\lambda(r){\in}\mathcal{A} \Rightarrow \emptyseq{\notin}\mathcal{L}(T)$
    
    \item $\lambda(r){\in}\{\rightarrow,\wedge\} \Rightarrow \emptyseq{\in}\mathcal{L}(T_1) \land \emptyseq{\in}\mathcal{L}(T_2) \Leftrightarrow \emptyseq{\in}\mathcal{L}(T)$
    
    \item $\lambda(r){\in}\times \Rightarrow \emptyseq{\in}\mathcal{L}(T_1) \lor \emptyseq{\in}\mathcal{L}(T_2) \Leftrightarrow \emptyseq{\in}\mathcal{L}(T)$
    
    \item $\lambda(r){=}\circlearrowleft \Rightarrow \emptyseq{\in}\mathcal{L}(T_1) \Leftrightarrow \emptyseq{\in}\mathcal{L}(T)$

\end{itemize}

\subsection{Interpretation of Process Tree Characteristics}
The decision where to split a trace and the assignment of sub-traces to subtrees is based on the four characteristics per subtree and the process tree operator. 
Thus, each subtree is a gray-box for the approximation approach since only limited information is available.
Subsequently, we explain how we interpret the subtree's characteristics and how we utilize them in the splitting/assigning decision.

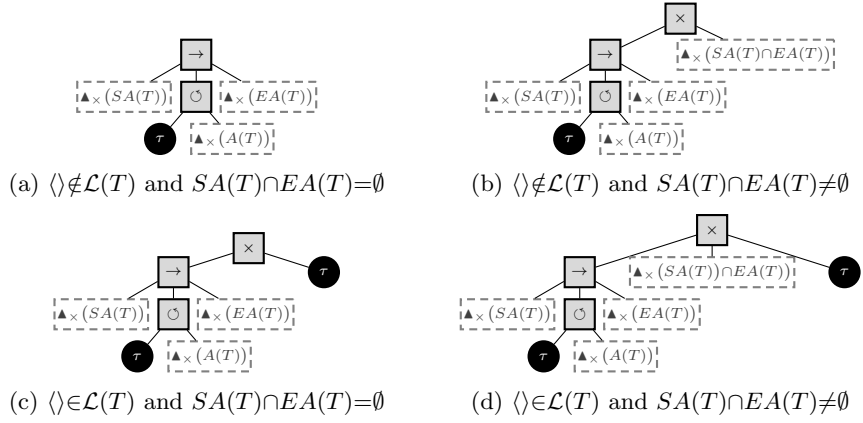
\begin{figure}[tb]
\begin{subfigure}[b]{.41\columnwidth}
    \scriptsize
    \centering
    \begin{tikzpicture}[every label/.style={text=darkgray,font=\scriptsize}, scale=0.8, every node/.style={scale=0.8}]
        \tikzset{
            tree_op/.style={rectangle,draw=black,fill=gray!30,thick,minimum size=6mm,inner sep=0pt},
            tree_subtree/.style={rectangle,draw = gray!15,fill=gray!15,thick,minimum size=6mm,inner sep=0pt},
            triangle/.style={inner sep=-1pt,draw=gray!30, regular polygon, regular polygon sides=3, align=center,equal size=T,fill=gray!30},
            marking/.style={draw=blue, densely dotted,thin,fill=blue!10},
            selected/.style={fill=darkgray, text=white, draw=gray},
            level 1/.style={sibling distance=13mm,level distance=14mm},
            level 2/.style={sibling distance=14mm},
            level 3/.style={sibling distance=20mm},
            level 4/.style={sibling distance=11mm,level distance=11mm},
            equal size/.style={execute at begin
             node={\setbox\nodebox=\hbox\bgroup},
             execute at end
             node={\egroup\eqmakebox[#1][c]{\copy\nodebox}}}
        }
    
        \tikzstyle{tree_op}=[rectangle,draw=black,fill=gray!30,thick,minimum size=5mm,inner sep=0pt]
        \tikzstyle{tree_leaf}=[circle,draw=black,thick,minimum size=5mm,inner sep=0pt]
        \tikzstyle{tree_leaf_inv}=[circle,draw=black,fill=black,thick,minimum size=5mm, text=white,inner sep=0pt]
        \tikzstyle{background}=[fill=black, fill opacity=.15, draw=lightgray]
        \tikzstyle{subtree}=[fill=white, text=darkgray, draw=gray, densely dashed, inner sep=1pt]

        \tikzstyle{level 1}=[sibling distance=12mm,level distance=7mm]
        \tikzstyle{level 2}=[sibling distance=12mm,level distance=7mm]
        \tikzstyle{level 3}=[sibling distance=12mm, level distance=7mm]
        \tikzstyle{level 4}=[sibling distance=10mm,level distance=9mm]
      
        \node [tree_op, label=] (root){$\rightarrow $}
                child {node [tree_op, subtree] (loop) {$\blacktriangle_\times\big(SA(T)\big)$}}
                child {node [tree_op, ] (parallel2) {$\circlearrowleft$}
                child {node [tree_leaf_inv,] (e) {$\tau$}}
                  child {node [tree_op, subtree] (loop) {$\blacktriangle_\times\big(A(T)\big)$}}
                }
                child {node [tree_op, subtree] (parallel2) {$\blacktriangle_\times\big(EA(T)\big)$}}
            
        ;
       
    \end{tikzpicture}
    \caption{$\emptyseq{\notin}\mathcal{L}(T)$ and $SA(T) {\cap} EA(T){=}\emptyset$}
    \label{fig:interpretation_1}
\end{subfigure}
\begin{subfigure}[b]{.58\columnwidth}
    \scriptsize
    \centering
    
    \begin{tikzpicture}[every label/.style={text=darkgray,font=\scriptsize},scale=0.8, every node/.style={scale=0.8}]
        \tikzset{
            tree_op/.style={rectangle,draw=black,fill=gray!30,thick,minimum size=6mm,inner sep=0pt},
            tree_subtree/.style={rectangle,draw = gray!15,fill=gray!15,thick,minimum size=6mm,inner sep=0pt},
            triangle/.style={inner sep=-1pt,draw=gray!30, regular polygon, regular polygon sides=3, align=center,equal size=T,fill=gray!30},
            marking/.style={draw=blue, densely dotted,thin,fill=blue!10},
            selected/.style={fill=darkgray, text=white, draw=gray},
            level 1/.style={sibling distance=12mm,level distance=7mm},
            level 2/.style={sibling distance=12mm},
            level 3/.style={sibling distance=20mm},
            level 4/.style={sibling distance=11mm,level distance=11mm},
            equal size/.style={execute at begin
             node={\setbox\nodebox=\hbox\bgroup},
             execute at end
             node={\egroup\eqmakebox[#1][c]{\copy\nodebox}}}
        }
    
        \tikzstyle{tree_op}=[rectangle,draw=black,fill=gray!30,thick,minimum size=5mm,inner sep=0pt]
        \tikzstyle{tree_leaf}=[circle,draw=black,thick,minimum size=5mm,inner sep=0pt]
        \tikzstyle{tree_leaf_inv}=[circle,draw=black,fill=black,thick,minimum size=5mm, text=white,inner sep=0pt]
        \tikzstyle{background}=[fill=black, fill opacity=.15, draw=lightgray]
        \tikzstyle{subtree}=[fill=white, text=darkgray, draw=gray, densely dashed, inner sep=1pt]

        \tikzstyle{level 1}=[sibling distance=25mm,level distance=6mm]
        \tikzstyle{level 2}=[sibling distance=12mm,level distance=7mm]
        \tikzstyle{level 3}=[sibling distance=12mm, level distance=7mm]
        \tikzstyle{level 4}=[sibling distance=10mm,level distance=9mm]
      
        \node [tree_op, label=] (root){$\times $}
            child {node [tree_op,label=] (seq) {$\rightarrow$}
                child {node [tree_op, subtree] (loop) {$\blacktriangle_\times\big(SA(T)\big)$}}
                child {node [tree_op, ] (parallel2) {$\circlearrowleft$}
                child {node [tree_leaf_inv,] (e) {$\tau$}}
                  child {node [tree_op, subtree] (loop) {$\blacktriangle_\times\big(A(T)\big)$}}
                }
                child {node [tree_op, subtree] (parallel2) {$\blacktriangle_\times\big(EA(T)\big)$}}
            }
            child {node [tree_op, subtree] (inter) {$\blacktriangle_\times\big(SA(T) {\cap} EA(T)\big)$}}
        ;
       
    \end{tikzpicture}
    \caption{$\emptyseq{\notin}\mathcal{L}(T)$ and $SA(T) {\cap} EA(T){\neq}\emptyset$}

    \label{fig:interpretation_2}
\end{subfigure}
\begin{subfigure}[b]{.41\columnwidth}
    \scriptsize
    \centering
    
    \begin{tikzpicture}[every label/.style={text=darkgray,font=\scriptsize},scale=0.8, every node/.style={scale=0.8}]
        \tikzset{
            tree_op/.style={rectangle,draw=black,fill=gray!30,thick,minimum size=6mm,inner sep=0pt},
            tree_subtree/.style={rectangle,draw = gray!15,fill=gray!15,thick,minimum size=6mm,inner sep=0pt},
            triangle/.style={inner sep=-1pt,draw=gray!30, regular polygon, regular polygon sides=3, align=center,equal size=T,fill=gray!30},
            marking/.style={draw=blue, densely dotted,thin,fill=blue!10},
            selected/.style={fill=darkgray, text=white, draw=gray},
            level 1/.style={sibling distance=13mm,level distance=14mm},
            level 2/.style={sibling distance=14mm},
            level 3/.style={sibling distance=20mm},
            level 4/.style={sibling distance=11mm,level distance=11mm},
            equal size/.style={execute at begin
             node={\setbox\nodebox=\hbox\bgroup},
             execute at end
             node={\egroup\eqmakebox[#1][c]{\copy\nodebox}}}
        }
    
        \tikzstyle{tree_op}=[rectangle,draw=black,fill=gray!30,thick,minimum size=5mm,inner sep=0pt]
        \tikzstyle{tree_leaf}=[circle,draw=black,thick,minimum size=5mm,inner sep=0pt]
        \tikzstyle{tree_leaf_inv}=[circle,draw=black,fill=black,thick,minimum size=5mm, text=white,inner sep=0pt]
        \tikzstyle{background}=[fill=black, fill opacity=.15, draw=lightgray]
        \tikzstyle{subtree}=[fill=white, text=darkgray, draw=gray, densely dashed, inner sep=1pt]

        \tikzstyle{level 1}=[sibling distance=25mm,level distance=4mm]
        \tikzstyle{level 2}=[sibling distance=12mm,level distance=7mm]
        \tikzstyle{level 3}=[sibling distance=12mm, level distance=7mm]
        \tikzstyle{level 4}=[sibling distance=10mm,level distance=9mm]
      
        \node [tree_op, label=] (root){$\times $}
            child {node [tree_op,label=] (seq) {$\rightarrow$}
                child {node [tree_op, subtree] (loop) {$\blacktriangle_\times\big(SA(T)\big)$}}
                child {node [tree_op, ] (parallel2) {$\circlearrowleft$}
                child {node [tree_leaf_inv,] (e) {$\tau$}}
                  child {node [tree_op, subtree] (loop) {$\blacktriangle_\times\big(A(T)\big)$}}
                }
                child {node [tree_op, subtree] (parallel2) {$\blacktriangle_\times\big(EA(T)\big)$}}
            }
            child {node [tree_leaf_inv,label=] (tau) {$\tau$}}
        ;
       
    \end{tikzpicture}
    \caption{$\emptyseq{\in}\mathcal{L}(T)$ and $SA(T) {\cap} EA(T){=}\emptyset$}

    \label{fig:interpretation_3}
\end{subfigure}
\hfill
\begin{subfigure}[b]{.58\columnwidth}
    \scriptsize
    \centering
    \begin{tikzpicture}[every label/.style={text=darkgray,font=\scriptsize}, scale=0.8, every node/.style={scale=0.8}]
        \tikzset{
            tree_op/.style={rectangle,draw=black,fill=gray!30,thick,minimum size=6mm,inner sep=0pt},
            tree_subtree/.style={rectangle,draw = gray!15,fill=gray!15,thick,minimum size=6mm,inner sep=0pt},
            triangle/.style={inner sep=-1pt,draw=gray!30, regular polygon, regular polygon sides=3, align=center,equal size=T,fill=gray!30},
            marking/.style={draw=blue, densely dotted,thin,fill=blue!10},
            selected/.style={fill=darkgray, text=white, draw=gray},
            level 1/.style={sibling distance=10mm,level distance=14mm},
            level 2/.style={sibling distance=14mm},
            level 3/.style={sibling distance=20mm},
            level 4/.style={sibling distance=11mm,level distance=11mm},
            equal size/.style={execute at begin
             node={\setbox\nodebox=\hbox\bgroup},
             execute at end
             node={\egroup\eqmakebox[#1][c]{\copy\nodebox}}}
        }
    
        \tikzstyle{tree_op}=[rectangle,draw=black,fill=gray!30,thick,minimum size=5mm,inner sep=0pt]
        \tikzstyle{tree_leaf}=[circle,draw=black,thick,minimum size=5mm,inner sep=0pt]
        \tikzstyle{tree_leaf_inv}=[circle,draw=black,fill=black,thick,minimum size=5mm, text=white,inner sep=0pt]
        \tikzstyle{background}=[fill=black, fill opacity=.15, draw=lightgray]
        \tikzstyle{subtree}=[fill=white, text=darkgray, draw=gray, densely dashed, inner sep=1pt]

        \tikzstyle{level 1}=[sibling distance=22mm,level distance=7mm]
        \tikzstyle{level 2}=[sibling distance=12mm,level distance=7mm]
        \tikzstyle{level 3}=[sibling distance=12mm, level distance=7mm]
        \tikzstyle{level 4}=[sibling distance=10mm,level distance=9mm]
      
        \node [tree_op, label=] (root){$\times $}
            child {node [tree_op,label=] (seq) {$\rightarrow$}
                child {node [tree_op, subtree] (loop) {$\blacktriangle_\times\big(SA(T)\big)$}}
                child {node [tree_op, ] (parallel2) {$\circlearrowleft$}
                child {node [tree_leaf_inv,] (e) {$\tau$}}
                  child {node [tree_op, subtree] (loop) {$\blacktriangle_\times\big(A(T)\big)$}}
                }
                child {node [tree_op, subtree] (parallel2) {$\blacktriangle_\times\big(EA(T)\big)$}}
            }
            child {node [tree_op, subtree,label={[xshift=-.3cm, yshift=0cm]}] (inter) {$\blacktriangle_\times\big(SA(T)\big) {\cap} EA(T)\big)$}}
                        child {node [tree_leaf_inv,] (tau) {$\tau$}}
        ;
       
    \end{tikzpicture}
        \caption{$\emptyseq{\in}\mathcal{L}(T)$ and $SA(T) {\cap} EA(T){\neq}\emptyset$}

    \label{fig:interpretation_4}
\end{subfigure}

\caption{Most liberal interpretation $\mathcal{I}(T)$ of the four characteristics of a process tree $T{\in}\mathcal{T}$. For a set $X{=}\{x_1,\dots,x_n\}$, $\blacktriangle_\times(X)$ represents the tree $\times(x_1,\dots,x_n)$}
\label{fig:interpretation}
\end{figure}

Consider \autoref{fig:interpretation} showing how the approximation approach assumes a given subtree $T$ behaves based on its four characteristics, i.e., $A(T),\allowbreak SA(T),\allowbreak EA(T),\allowbreak \emptyseq{\in}\mathcal{L}(T)$.
The most liberal \emph{interpretation} $\mathcal{I}(T)$ of a subtree $T$ can be considered as a heuristic that guides the splitting/assigning.
The interpretation $\mathcal{I}(T)$ depends on two conditions, i.e., if $\emptyseq{\in}\mathcal{L}(T)$ and whether there is an activity that is both, a start- and end-activity, i.e., $SA(T) {\cap} EA(T){\neq}\emptyset$.
Note that $\mathcal{L}(T) {\subseteq} \mathcal{L}(\mathcal{I}(T))$ holds.
Thus, the interpretation is an approximated view on the actual subtree.

In the next sections, we present for each tree operator a splitting/assigning and composing strategy based on the presented subtree interpretation.
All strategies return a splitting per recursive call that minimizes the overall edit distance between the sub-traces and the closest trace in the language of the interpretation of the assigned subtrees.  
For $\sigma_1,\sigma_2 {\in}\mathcal{A}^*$, let $\mathcal{l}(\sigma_1,\sigma_2){\in}\mathbb{N}{\cup}\{0\}$ be the Levenshtein distance~\cite{levenshtein1966binary}.
For given $\sigma{\in}\mathcal{A}^*$ and $T{\in}\mathcal{T}$, we calculate a valid splitting $\psi(\sigma,T){=}\allowbreak\big\langle (\sigma_1,T_{i_1}),\allowbreak \dots, \allowbreak (\sigma_j,T_{i_k})\big\rangle$ w.r.t. \cref{eq:splitting} s.t. the sum depicted below is minimal.
\begin{equation}
\label{eq:splitting_interpretation}
    \sum_{j{\in}\{1,\dots,k\}}\big(\min\limits_{\sigma' {\in} \mathcal{I}(T_{i_j})} \mathcal{l}(\sigma_j,\sigma')\big)
\end{equation}

In the upcoming sections, we assume a given trace $\sigma{=}\langle a_1,\allowbreak\dots,\allowbreak a_n\rangle$ and a process tree $\pt$ with subtrees referred to as $T_1$ and $T_2$.


\subsection{Approximating on Choice Operator}

The choice operator is the most simple one since we just need to assign $\sigma$ to one of the subtrees according to the semantics, i.e., assigning $\sigma$ either to $T_1$ or $T_2$.
We compute the edit distance of $\sigma$ to the closest trace in $\mathcal{I}(T_1)$ and in $\mathcal{I}(T_2)$ and assign $\sigma$ to the subtree with smallest edit distance according to \cref{eq:splitting_interpretation}.

Composing an alignment for the choice operator is trivial.
Assume we eventually get an alignment $\gamma$ for the chosen subtree, we just return $\gamma$ for $T$.

\subsection{Approximating on Sequence Operator}
When splitting on a sequence operator, we must assign a sub-trace to each subtree according to the semantics.
Hence, we calculate two sub-traces: $\langle(\sigma_1,T_1),\allowbreak (\sigma_2,T_2)\rangle$ s.t. $\sigma_1 {\cdot} \sigma_2 {=} \sigma$ according to \cref{eq:splitting_interpretation}.
The optimal splitting/assigning can be defined as an optimization problem, i.e., Integer Linear Programming (ILP).

In general, for a trace with length $n$, $n{+}1$ possible splitting-positions exist:
$
\langle \textcolor{gray}{|_1}\ a_1 \ \textcolor{gray}{|_2} \ a_2 \ \textcolor{gray}{|_3} \ \dots\allowbreak \ \textcolor{gray}{|_n} \ a_n \ \textcolor{gray}{|_{n{+}1}} \rangle
$.
Assume we split at position $1$, this results in $\big\langle (\emptyseq,T_1), \allowbreak(\sigma,T_2) \big\rangle$, i.e., we assign $\emptyseq$ to $T_1$ and the original trace $\sigma$ to $T_2$. 

Composing the alignment from sub-alignments is straightforward. 
In general, we eventually obtain two alignments, i.e, $\langle\gamma_1,\gamma_2\rangle$, for $T_1$ and $T_2$. 
We compose the alignment $\gamma$ for $T$ by concatenating the sub-alignments, i.e., $\gamma{=} \gamma_1{\cdot}\gamma_2$.

\subsection{Approximating on Parallel Operator}
\label{sec:approx_on_parallel_operator}
According to the semantics, we must define a sub-trace for each subtree, i.e., $\langle (T_1,\sigma_1),\allowbreak (T_2,\sigma_2) \rangle$.
In contrast to the sequence operator, $\sigma_1{\cdot}\sigma_2{=}\sigma$ does \emph{not} generally hold. 
The splitting/assignment w.r.t. \cref{eq:splitting_interpretation} can be defined as an ILP.
In general, each activity can be assigned to one of the subtrees independently. 

For example, assume $\sigma {=} \langle c,a,d,c,b \rangle$ and $T\widehat{=} {\wedge}\big({\rightarrow}(a,b),\allowbreak{\circlearrowleft}(c,d)\big)$ with subtree $T_1\widehat{=} {\rightarrow}(a,b)$ and $T_2\widehat{=} {\circlearrowleft}(c,d)$.
Below we assign the activities to subtrees.
\vspace{-3mm}
\begin{center}
\bgroup
\setlength\tabcolsep{1.5mm}
\begin{tabular}{cccccccccc}
$\langle$ & $c$,          & $a$,               & $d$,          & $c$,          & $b$ & $\rangle$\\
        & $T_2$  & $T_1$       & $T_2$          & $T_2$         & $T_1$
\end{tabular}
\egroup
\end{center}
\vspace{-3mm}
Based on the assignment, we create two sub-traces: $\sigma_1{=}\langle a,b\rangle$ and $\sigma_2{=}\langle c,d,c \rangle$.
Assume that $\gamma_1{\widehat{=}}\langle (a,a),(b,b)\rangle$ and $\gamma_2{\widehat{=} }\langle (c,c),(d,d),(c,c)\rangle$ are the two alignments eventually obtained.
To compose an alignment for $T$, we have to consider the assignment.
Since the first activity $c$ is assigned to $T_2$, we extract the corresponding alignment steps from $\gamma_1$ until we have explained $c$.
The next activity in $\sigma$ is an $a$ assigned to $T_1$.
We extract the alignment moves from $\gamma_1$ until we explained the $a$.
We iteratively continue until all activities in $\sigma$ are covered.
Finally, we obtain an alignment for $T$ and $\sigma$, i.e., $\gamma{\widehat{=} } \big\langle (c,c),(a,a),(d,d),(c,c),(b,b) \big\rangle$.

\subsection{Approximating on Loop Operator}

We calculate $m{\in}\{1,3,5,\dots\}$ sub-traces that are assigned alternately to the two subtrees: $\langle (\sigma_1,T_1),(\sigma_2,T_2),(\sigma_3,T_1),\dots,(\sigma_{m-1},T_2),(\sigma_m,T_1) \rangle$ s.t. $\sigma{=}\sigma_1 {\cdot} \dots{\cdot} \sigma_m$.
Thereby, $\sigma_1$ and $\sigma_m$ are always assigned to $T_1$.
Next, we visualize all possible splitting positions for the given trace:
$
\langle \textcolor{gray}{\mid_1} \:a_1 \:\textcolor{gray}{\mid_2}\:\textcolor{gray}{\mid_3} \:   a_2\:\textcolor{gray}{\mid_4}   \dots   \textcolor{gray}{\mid_{2n{-}1}} \:a_n \:\textcolor{gray}{\mid_{2n}}   \rangle
$.
If we split at each position, we obtain $\big\langle \big( \emptyseq,T_1\big), \big(\langle a_1\rangle,T_2\big),\big(\emptyseq,T_1\big),\allowbreak \dots,\allowbreak \big(\langle a_n\rangle,T_2\big), \big(\emptyseq,T_1\big) \big\rangle $.
The optimal splitting/assignment w.r.t \cref{eq:splitting_interpretation} can be defined as an ILP.

Composing an alignment is similar to the sequence operator. 
In general, we obtain $m$ sub-alignments $\langle \gamma_1,\dots,\gamma_m \rangle$, which we concatenate, i.e., $\gamma {=}\gamma_1 {\cdot}\dots{\cdot} \gamma_m$. 

\section{Evaluation}
\label{sec:evaluation}

This section presents an experimental evaluation of the proposed approach.


We implemented the proposed approach in PM4Py\footnote{\url{https://pm4py.fit.fraunhofer.de/}}, an open-source process mining library.
We conducted experiments on real event logs \cite{bpi_ch_19,bpi_ch_18}.
For each log, we discovered a process tree with the Inductive Miner infrequent algorithm~\cite{DBLP:conf/bpm/LeemansFA13}.

\begin{figure}[tb]
        \begin{subfigure}[t]{.46\columnwidth}
        \centering
        \includegraphics[width=\columnwidth,trim=.5cm .4cm 0.7cm 0.3cm,clip]{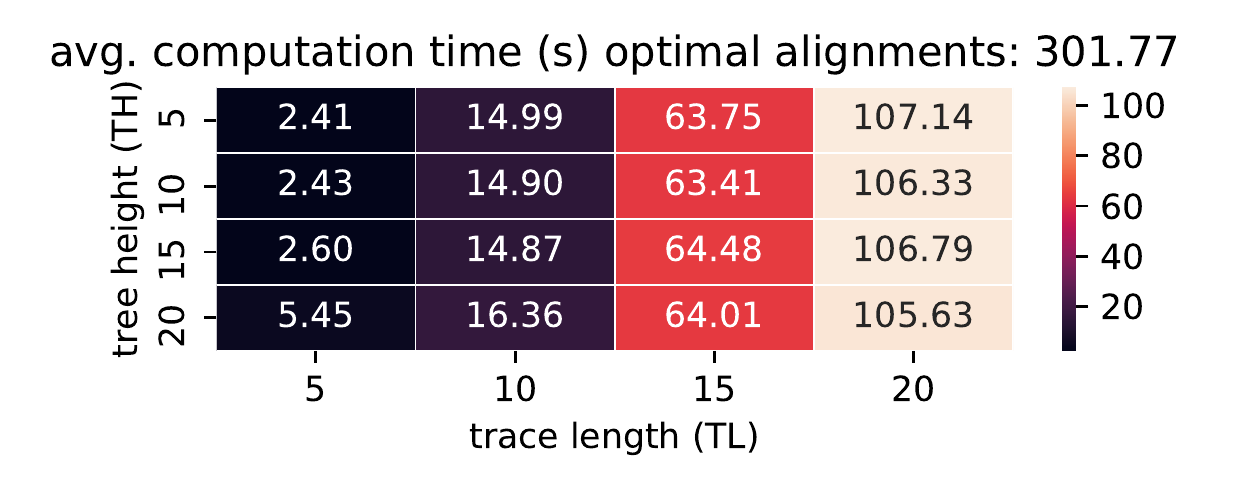}
        \caption{Avg. computation time (s)}
        \label{fig:alignment_duration_heatmap_bpi_ch_19}
    \end{subfigure}
    \hfill
    \begin{subfigure}[t]{.46\columnwidth}
        \centering
        \includegraphics[width=\columnwidth,trim=.4cm .4cm 0.6cm 0cm,clip]{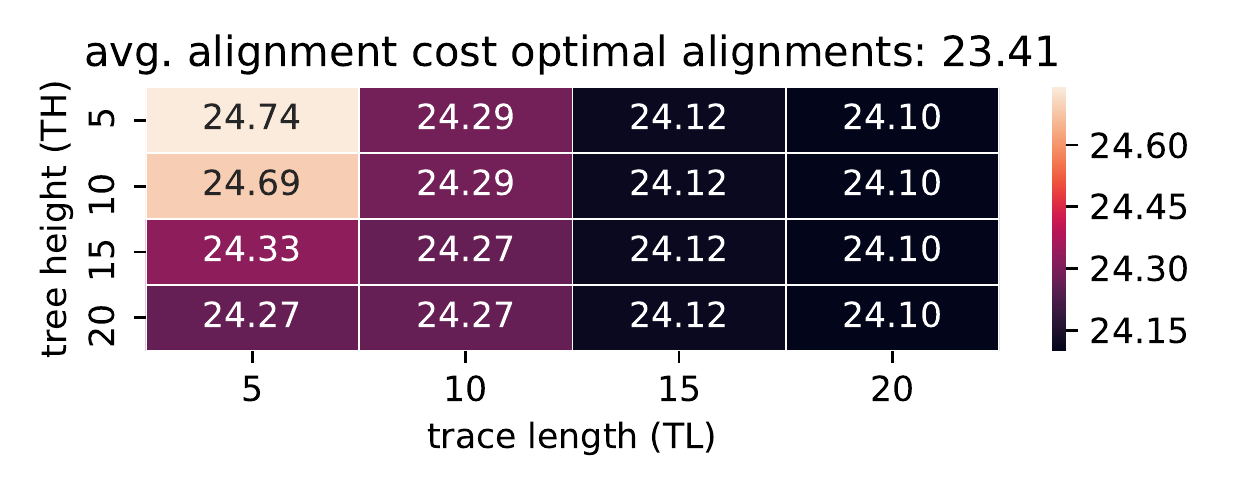}
        \caption{Avg. alignment costs}
        \label{fig:alignment_cost_heatmap_bpi_ch_19}
    \end{subfigure}
    \caption{Results for~\cite{bpi_ch_19}, sample: 100 variants, tree height $24$, avg. trace length $28$}
    \label{fig:results_bpi_ch_19}
\end{figure}

\begin{figure}[tb]
        \begin{subfigure}[t]{.46\columnwidth}
        \centering
        \includegraphics[width=\columnwidth,trim=.5cm .4cm 0.7cm 0.3cm,clip]{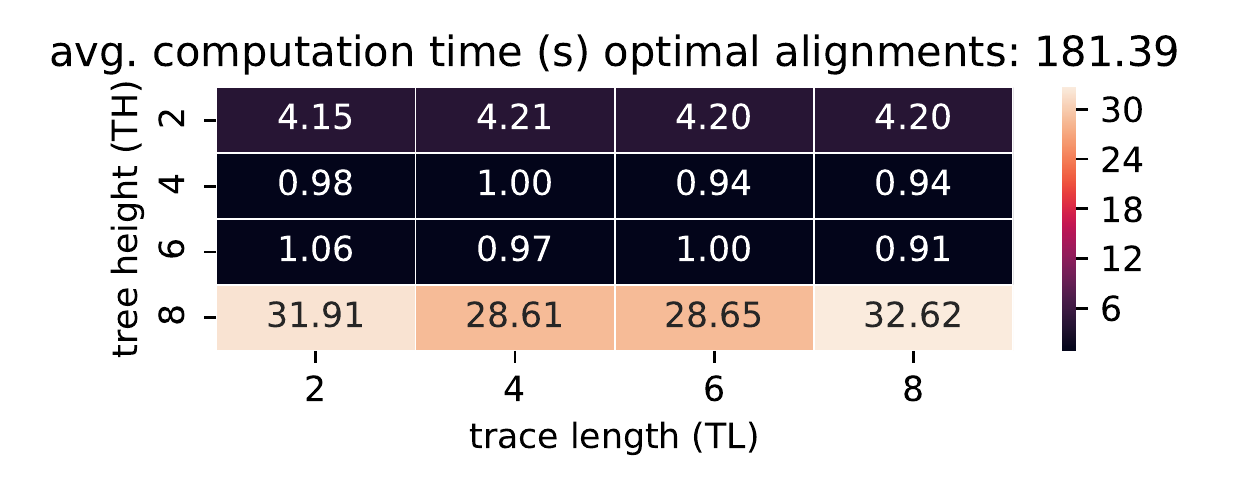}
        \caption{Avg. computation time (s)}
        \label{fig:alignment_duration_heatmap_bpi_ch_18}
    \end{subfigure}
    \hfill
    \begin{subfigure}[t]{.46\columnwidth}
        \centering
        \includegraphics[width=\columnwidth,trim=.4cm .4cm 0.6cm 0cm,clip]{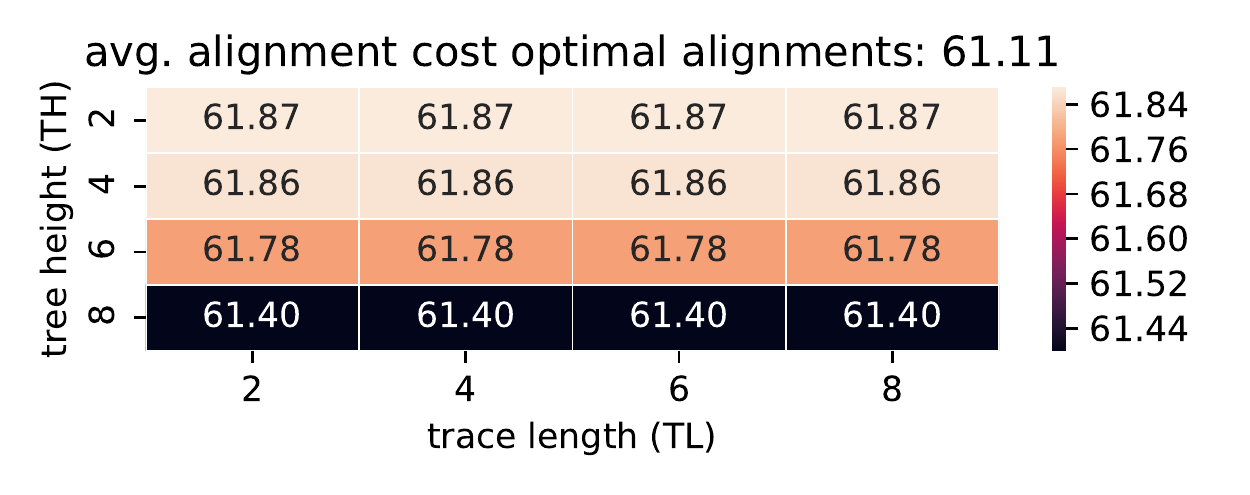}
        \caption{Avg. alignment costs}
        \label{fig:alignment_cost_heatmap_bpi_ch_18}
    \end{subfigure}
    \caption{Results for \cite{bpi_ch_18}, sample: 100 variants, tree height $10$, avg. trace length $65$}
    \label{fig:results_bpi_ch_18}
\end{figure}

\begin{table}[tb]
\caption{Results for decomposition based alignments}
\label{tab:decomposition}
\centering
\scriptsize
\begin{tabular}{|c||c|c|}
\hline
Approach                              & \cite{bpi_ch_19} (sample: 100 variants) & \cite{bpi_ch_18} (sample: 100 variants)\\ \hline\hline
decomposition \cite{DBLP:journals/isci/LeeVMAS18} & 25.22 s      &    20.96 s        \\ \hline
standard \cite{adriansyah_2014_phd_aligning}   & 1.51 s      &       103.22 s    \\ \hline
\end{tabular}
\end{table}

In Figures \ref{fig:results_bpi_ch_19} and \ref{fig:results_bpi_ch_18}, we present the results.
We observe that our approach is on average always faster than the optimal alignment algorithm for all tested parameter settings.
Moreover, we observe that our approach never underestimates the optimal alignment costs, as our approach returns a valid alignment. 
W.r.t. optimization problems for optimal splittings/assignments, consider parameter setting TH:5 and TL:5 in~\autoref{fig:results_bpi_ch_19}.
This parameter setting results in the highest splitting along the tree hierarchy and the computation time is the lowest compared to the other settings.
Thus, we conclude that solving optimization problems for finding splittings/assignments is appropriate.
In general, we observe a good balance between accuracy and computation time.
We additionally conducted experiments with a decomposition approach~\cite{DBLP:journals/dpd/Aalst13} (available in ProM\footnote{\url{http://www.promtools.org/}}) and compared the calculation time with the standard alignment implementation (LP-based)~\cite{adriansyah_2014_phd_aligning} in ProM.
Consider \autoref{tab:decomposition}.
We observe that the decomposition approach does not yield a speed-up for \cite{bpi_ch_19} but for \cite{bpi_ch_18} we observe that the decomposition approach is about 5 times faster. 
In comparison to \autoref{fig:alignment_duration_heatmap_bpi_ch_18}, however, our approach yields a much higher speed-up.

\section{Conclusion}
\label{sec:conclusion}
We introduced a novel approach to approximate alignments for process trees.
First, we recursively split a trace into sub-traces along the tree hierarchy based on a gray-box view on the respective subtrees.
After splitting, we compute optimal sub-alignments.
Finally, we recursively compose a valid alignment from sub-alignments.
Our experiments show that the approach provides a good balance between accuracy and calculation time.
Apart from the specific approach proposed, the contribution of this paper is the formal framework describing how alignments can be approximated for process trees.
Thus, many other strategies besides the one presented are conceivable.

\bibliographystyle{IEEEtran}
\bibliography{main}

\end{document}